\documentclass[11pt]{article}
\usepackage{xspace,amssymb,amsmath,helvet,graphicx}
\begin{document}
\newcommand{\dee}{\,\mbox{d}}
\newcommand{\naive}{na\"{\i}ve }
\newcommand{\eg}{e.g.\xspace}
\newcommand{\ie}{i.e.\xspace}
\newcommand{\pdf}{pdf.\xspace}
\newcommand{\etc}{etc.\@\xspace}
\newcommand{\PhD}{Ph.D.\xspace}
\newcommand{\MSc}{M.Sc.\xspace}
\newcommand{\BA}{B.A.\xspace}
\newcommand{\MA}{M.A.\xspace}
\newcommand{\role}{r\^{o}le}
\newcommand{\signoff}{\hspace*{\fill} Rose Baker \today}
\newenvironment{entry}[1]%
{\begin{list}{}{\renewcommand{\makelabel}[1]{\textsf{##1:}\hfil}%
\settowidth{\labelwidth}{\textsf{#1:}}%
\setlength{\leftmargin}{\labelwidth}
\addtolength{\leftmargin}{\labelsep}
\setlength{\itemindent}{0pt}
}}%
{\end{list}}
\title{Mathematical models of confirmation bias}
\author{Rose Baker\\School of Business\\University of Salford, UK}
\maketitle
\begin{abstract}
Confirmation bias is a cognitive bias that adversely affects management decisions, and mathematical modelling is an aid to its detailed understanding.
Bias in opinion update about the value of a parameter is modelled here assuming that observations are discounted 
depending on their distance from prior opinion. The models allow belief persistence, attitude polarization, and the irrational primacy effect to be explored.
A general framework for exploring large-sample properties of these models is given, and an attempt made to classify the models.
An interesting result is that in some models the influence of an observation always increases with distance from the prior opinion, whereas in others observations
greatly at odds with prior opinion are given very little weight.
The models could be useful to those exploring these phenomena in detail.
\end{abstract}
\section*{Keywords}
Decision theory; cognitive bias; confirmation bias; log-gamma distribution; directional discounting.
\section{Introduction}
\subsection{Managerial relevance of confirmation bias and its modelling}
Decision-making is essential to management: French (1991, 1995) gives a description.
However, management decisions, like all others,  are plagued by a vast host of cognitive biases. Sutherland (2013) gives a comprehensive account (the first edition of his
work dates from 1992).
Among cognitive biases, confirmation bias, the tendency to interpret data so as to confirm one's prior beliefs, is arguably the chief. 
It has been quoted as causing belief perseverance, attitude polarization, and the irrational primacy effect (earlier observations have more impact).
 
An example is the launch of a new product line. The CEO initially `knows' it is a good idea, and the subsequent feasibility research is interpreted so as to support
this belief, rendering the research valueless. Another is the recruitment of new staff, where a battery of aptitude tests cannot dislodge the hirers' original
impression of the potential new recruit, impressions perhaps based on unconscious bias.

The relevance of management studying and becoming knowledgeable about confirmation bias is evident.
In general, mathematical modelling can be the path to more detailed understanding of phenomena,
such as has been the case with  infectious diseases. Models can offer detailed predictions, but more importantly they can give insight and help to focus further study.
Here, however, little mathematical modelling has been done.
\subsection{Modelling confirmation bias}
Previous work includes Rabin and Schrag (1999), who considered a series of Bernoulli trials, where signals that conflict with current belief are sometimes misinterpreted.
Gerber and Green (1999) considered biased learning, \eg normally-distributed observations, where the random variable $X$ is taken as $\alpha X$ if above a neutral position at zero.
This comes close to the ideas presented here.
Zimper and Ludwig (2009) develop models of Bayesian learning using Choquet expected utility theory, and Jern {\em et al} (2014) considered belief polarization arising from Bayesian networks.
Allahverdyan and Galstyan (2014) considered a non-Bayesian model. This work extends Baker (2021).

Here, we discuss some simple Bayesian models of the evolution of belief about observable quantities, such as an individual's blood pressure, or an organization's sales figures.
In these models bias is caused by undue discounting of observations that contradict prior belief.

We first consider rational updating of opinion in the light of observation, where Bayes' theorem tells us how we ought to update our prior belief.

Let the probability density function (pdf) of prior belief about a parameter $\theta$ be $f(\theta)$, and the likelihood (probability or pdf) of an observation $X$ be ${\cal L}(X|\theta)$.
The the posterior pdf of the distribution of $\theta$ is 
\begin{equation}f(\theta|X)=\frac{{\cal L}(X|\theta)f(\theta)}{\int {\cal L}(X|\phi)f(\phi)\dee \phi}.\label{eq:bayes}\end{equation}
In this work, the likelihood is that for a normal distribution with mean $\mu$ and variance $\sigma^2$.
Unless the prior belief $f$ is certainty, after $n$ observations the posterior distribution will tend to a narrow Gaussian centred on $\mu$ with variance approximately $\sigma^2/n$.

Bayesian meta-analysis provides an example of this kind in which opinion is updated in the light of a series of observations (studies).
In general, the problem with using (\ref{eq:bayes}) is that we do not really know $\sigma^2$.
The problem is thus with the likelihood function rather than the prior belief.
In general, we must interpret the world in the light of our current (prior) knowledge, and in a world of fake news and hype, claimed results must often be discounted to some extent.
Even in meta-analysis, where a series of results with well-estimated variances is available, in over half of analyses the true error is found to exceed the quoted one,
as judged by the scatter of the results. 
In medicine this may be caused by different patient mixes or different operating procedures, but the same problem arises in other contexts, such as physics.
Thus discounting data is not irrational, although it can lead to the persistence of prior belief 
even given a lot of contrary evidence. 

There is a great deal of irrational and bizarre belief in the world,
but this work is concerned with individuals and organizations that are at least trying to draw correct conclusions from data, but may still fall victim to confirmation bias.

The next section introduces a simple model of bias, then a general framework for analysing models is given. Finally there is an attempt to classify models
and a discussion of primacy, followed by some conclusions.

\section{Models of bias}
Imagine a series of $n$ independent observations $X_i$, where $X_i \sim N[\mu,\sigma^2]$.  For example, the random variable could be 
blood pressure or some other medical measurement for an individual, or a quantity of corporate interest such as a sales figure.
There are more complex cases where variances $\sigma_i^2$ differ, or where $\mu$ or $\sigma^2$ changes with time, but for simplicity these are not considered.

We wish to find a posterior distribution for the centre of location. Using (\ref{eq:bayes}) starting from a normally-distributed prior with mean $\mu_0$, variance $\sigma_0^2$,
after $n$ observations the posterior distribution would be normal, with variance $v$ given by $1/v=n/\sigma^2+1/\sigma_0^2$, mean $(n\bar{X}/\sigma^2+\mu_0/\sigma_0^2)/v$.
Asymptotically as $n\rightarrow\infty$ we can ignore prior belief.
However, with biased observation the posterior distribution will be different.

Denote the centre of location by $\theta$ and the mean of the posterior distribution of $\theta$ by $\lambda$

In these models, observations are discounted preferentially in one direction, which to coin a phrase we could call `directional discounting'. 
Intuitively, we can see that results in the `wrong' direction might be less-liked. Utility theory and prospect theory (\eg Kahneman, 2012)
teach that the pain of losing is greater than the satisfaction of an equivalent gain.

The discounting has here been taken as `low is good', but the models can easily be changed to the opposite.
The true distribution of observations is normal, with the mean taken as zero without loss of generality.
Hence here the mean of the posterior distribution is also the asymptotic bias $\lambda$.

Several models are explored, chosen for realism and relative ease of computation of the bias $\lambda$. 
The asymptotic properties are explored as $n \rightarrow \infty$. In real life there will often be only a few observations, but the asymptotic properties
will still shed light on how biased observation behaves.

Given a formula for $\hat\lambda$ after $n$ observations, a quantity of interest is the influence $\Delta\hat{\lambda}$ of an observation $X$. This is here defined as the change in $\hat\lambda$
following from adding a new $n+1$th observation $X$, when $n$ is very large. In the unbiased case, this increases linearly with $X-\lambda$.
It is convenient to use $n\Delta\hat{\lambda}$, which tends to a limit.

Some simple models are presented, followed by a `taxonomy' of models.
Computations were carried out via purpose-written Fortran95 programs, using the Numerical Algorithms Group (NAG) library for random-number generation and special function evaluation.

\subsection{Exponential Model}
\begin{figure}[h]
\centering
\makebox{\includegraphics{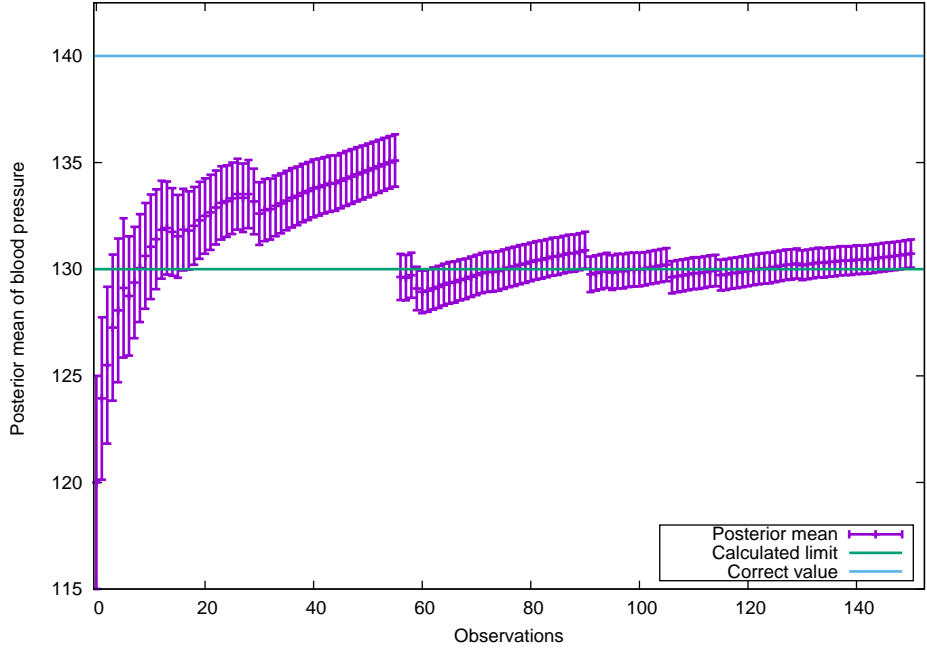}}
\caption{\label{fig1}Posterior mean and standard deviation of blood pressure, starting at a prior mean of 120 mm Hg, sd. 5, with random observations of
mean 140, sd. 10, and with exponential discounting, $\beta=0.2$.}
\end{figure}

The simplest model is that an individual discounts observations $X$ exponentially, so that the variance $\sigma^2$ is taken as $\sigma^2\exp(\beta X/\sigma)$.
These discounted observations update the prior opinion.
If $\beta > 0$, high observations are regarded as less trustworthy. 
This simple model is quite tractable but it ignores the fact that bias should be relative to current belief, \ie the variance should really be $\sigma^2\exp\{\beta(X-\theta)/\sigma\}$,
of which more later.
Figure \ref{fig1} shows the evolution of the mean and standard deviation of the posterior distribution under this model. The sudden downward jumps are caused by low observations, which have
a corresponding low perceived variance and large influence. The approach to the asymptotic limit can be seen.
The asymptotic distribution of belief will be normal, and the posterior distribution mean $\lambda$ and variance $v$ can be found after some large number $n$ of observations
using the expressions for a weighted mean
\begin{equation}\hat{\mu}=\frac{\sum_{i=1}^n X_i/\sigma_i^2}{\sum_{i=1}^n 1/\sigma_i^2}\label{eq:truemean}\end{equation}and its variance
\begin{equation}v=\frac{1}{\sum_{i=1}^n 1/\sigma_i^2}.\label{eq:truevar}\end{equation}
This gives here
\begin{equation}\hat{\lambda}=\frac{\sum_{i=1}^n X_i\exp(-\beta X_i/\sigma)}{\sum_{i=1}^n \exp(-\beta X_i/\sigma)}.\label{eq:expmean}\end{equation}
As $n \rightarrow\infty$ the numerator and denominator tend to $n \text{E}\{X\exp(-\beta X/\sigma)\}, \text{E}\{\exp(-\beta X/\sigma)\}$ respectively.
Using the results 
\begin{equation}\text{E}\{\exp(-\beta X/\sigma)\}=\frac{1}{\sqrt{2\pi\sigma^2}}\int_{-\infty}^\infty \exp(-x^2/2\sigma^2-\beta x/\sigma)\dee x=\exp(\beta^2/2),\label{eq:ident1}\end{equation}
\[\text{E}\{X\exp(-\beta X/\sigma)\}=\frac{1}{\sqrt{2\pi\sigma^2}}\int_{-\infty}^\infty x\exp(-x^2/2\sigma^2-\beta x/\sigma)\dee x=-\beta\sigma\exp(\beta^2/2),\]
we have that 
\begin{equation}\hat{\lambda}\rightarrow \lambda=-\beta\sigma.\label{eq:mean}\end{equation}
This shows that the mean of the posterior distribution is biased downwards if $\beta > 0$ and vice versa, by an amount that increases with
the standard deviation $\sigma$ of the observations. 

The perceived (subjective) estimate of the variance of $\hat{\lambda}$ from (\ref{eq:truevar}) is 
\[v=n\sigma^2/\text{E}\{\exp(-\beta X/\sigma)\}=(\sigma^2/n)\exp(-\beta^2/2),\]
which is smaller than the unbiased estimate of $\sigma^2/n$.
Since
\begin{equation}\hat{\lambda}-\lambda=\frac{\sum_{i=1}^n (X_i-\lambda)\exp(-\beta X_i/\sigma)}{\sum_{i=1}^n \exp(-\beta X_i/\sigma)}.\label{eq:diff}\end{equation}
the true variance of $\hat{\lambda}$ is given by
\[\text{var}(\hat{\lambda})=\frac{\text{E}\{\exp(-2\beta X/\sigma)(X -\lambda)^2\}}{n\text{E}\{\exp(-\beta X/\sigma)\}^2},\]
which evaluates to
\begin{equation}\text{var}(\hat{\lambda})=(1+\beta^2)\exp(\beta^2)(\sigma^2/n).\end{equation}
Whereas the subjective (perceived) variance was smaller than the unbiased variance, the true variance is larger, by a factor that increases rapidly with the bias $\beta$.

Finally, the distribution of $\hat\lambda$ is asymptotically normal, from (\ref{eq:diff}).
The numerator is a sum of random variates which is normally distributed by the Central Limit Theorem (CLT). The denominator does not alter this, as it tends to a constant.

Hence under this simple model, discounting high or low observations biases the mean of the posterior distribution by an amount that increases with the standard deviation.
In this case discounting reduces the subjective variance of the posterior distribution, although that does not always happen. It also increases the actual variance,
which does always happen.

\begin{figure}
\centering
\makebox{\includegraphics{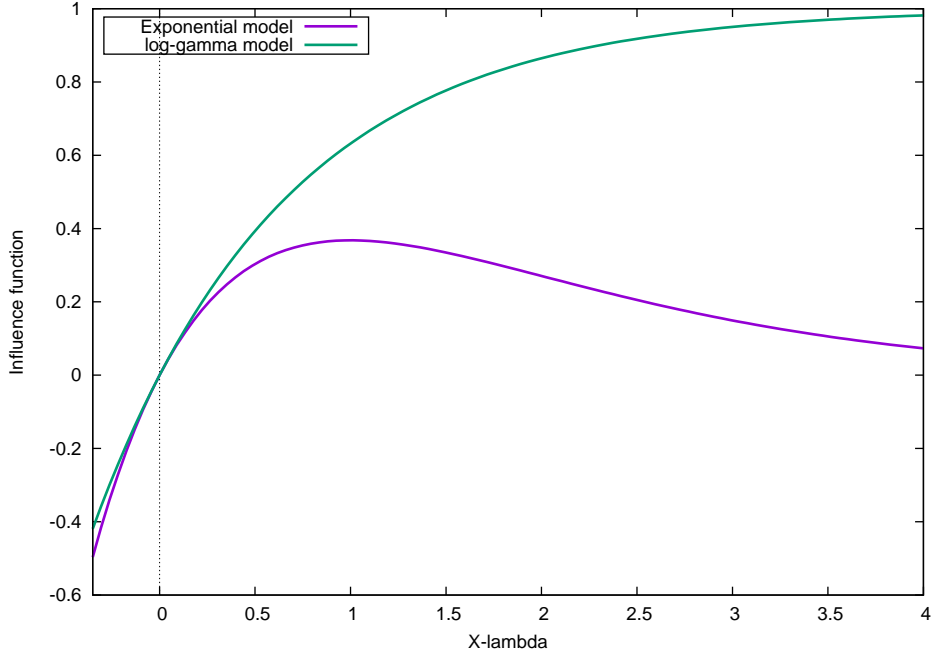}}
\caption{\label{fig2}The influence function $n\Delta\hat{\lambda}$ for the exponential model with $\beta=1$
and for the log-gamma distribution model, ditto.}
\end{figure}

The influence function is readily computed from (\ref{eq:expmean}) as

\begin{equation}n\Delta\hat{\lambda}=\exp\{-\beta(X/\sigma+\beta/2)\}(X-\lambda).\label{eq:counter}\end{equation}
This is plotted in figure \ref{fig2} for $\beta=\sigma=1$. For $X > \lambda$ it shows surprising behaviour. As $X$ increases, influence first increases as one might expect, but it then decreases after $X-\lambda=\sigma/\beta$,
\ie after $X > 0$. The discounting is so heavy that very high observations are effectively ignored. Paradoxically, stronger evidence is largely ignored as being too far from prior belief.

\subsubsection{Beta distribution model}
We can apply this type of exponential model when the observations are not normally distributed random variables such as sales figures, but are probabilities or percentages,
for example the efficacy of a vaccine expressed as a proportion rendered immune.
Then the discounting factor can be the odds ratio $\{(1-p)/p\}^\gamma=\exp\{\ln(\frac{1-p}{p})\gamma\}$, where higher probabilities are discounted.
The posterior distribution mean can be estimated as 
\[\hat{\lambda}=\frac{\sum_{i=1}^n p_i\{(1-p_i)/p_i\}^\gamma}{\sum_{i=1}^n \{(1-p_i)/p_i\}^\gamma},\]
which on assuming the $p_i$ distributed as a beta distribution with parameters $\alpha, \beta$ and taking expectations as $n \rightarrow\infty$ becomes 
\[B(\alpha+1-\gamma,\beta+\gamma)/B(\alpha-\gamma,\beta+\gamma)=\frac{\alpha-\gamma}{\alpha+\beta}=\mu-\frac{\gamma}{\alpha+\beta},\]
where $B$ denotes the beta function.
The variance of the posterior distribution is 
\[v=\frac{\alpha\beta}{(\alpha+\beta)^2(\alpha+\beta+1)}\frac{B(\alpha-\gamma,\beta+\gamma)}{B(\alpha,\beta)}\frac{1}{n}.\]
Note that we must have $\gamma < \alpha, \gamma > -\beta$.

\section{A likelihood-based approach}
The approach used for the exponential model and beta-distribution models cannot be used in general, 
where the bias is relative to the prior belief $\theta$. The simplest general way to derive the asymptotic behaviour of models is as follows.
For a specific prior belief $\theta$ we find the log-likelihood $\ell(X_1\cdots X_n|\theta)$. Asymptotically, after a large number $n$ of observations, this tends to its expected value
$\ell/n \rightarrow \text{E}\{\ell(X|\lambda)\}$ under the true distribution for $X \sim N[0,\sigma^2]$. Because the likelihood becomes strongly peaked,
the (subjective) variance of $\hat\lambda$ is given by 
\begin{equation}1/v=- \partial^2\text{E}(\ell)/\partial \lambda^2.\label{eq:v}\end{equation} 
This tends to zero as $n \rightarrow\infty$,
with the asymptotic value of $\lambda$ being that which maximises the likelihood, \ie we require
\begin{equation}\partial \text{E}(\ell)/\partial\lambda=0.\label{eq:lambda}\end{equation}

This approach yields the asymptotic bias $\lambda$ and the subjective (perceived) error. It also yields the the true variance $\text{var}(\hat{\lambda})$ 
of $\hat\lambda$.
This can be found by expanding the score function about $\lambda$:

\[\partial\ell/\partial\lambda|_{\hat{\lambda}}=0\simeq \partial\ell/\partial\lambda|_{\lambda}+\partial^2\ell/\partial\lambda^2|_{\lambda}(\hat{\lambda}-\lambda).\]
Hence 
\begin{equation}\hat{\lambda}-\lambda=-\frac{\partial\ell/\partial\lambda|_{\lambda}}{\partial^2\ell/\partial\lambda^2|_{\lambda} }\label{eq:approxlam}\end{equation}
The influence function can be found using this large-sample approximation for $\hat\lambda$. 

From (\ref{eq:approxlam})
\begin{equation}\text{var}\hat{\lambda}=\frac{\text{E}\{(\partial\ell/\partial\lambda)^2\}}{\text{E}(\partial^2\ell/\partial\lambda^2)^2}.\label{eq:var}\end{equation}
Normally the Bartlett identity $\text{E}(\partial^2\ell/\partial\theta^2)=-\text{E}\{(\partial\ell/\partial\theta)^2\}$ would mean that these two forms for $v$ and $\text{var}(\hat{\lambda})$ are the same,
but here this is not so because the likelihood function is wrongly specified, \ie the expectation is not taken over the subjective distribution.
The maximum-likelihood estimator tends to the posterior mean as $n\rightarrow\infty$, so the formulae above are the required variances.

With this framework, we can examine the `relative exponential' model, which puts right the unrealistic feature of the exponential bias model, that the weight given to an observation did not depend on the opinion.
Correcting this, with an opinion $\theta$ we obtain the log-likelihood
\[\ell=-\frac{\sum_{i=1}^n (X_i-\lambda)^2\exp(-\beta(X_i-\lambda)/\sigma)}{2\sigma^2}.\]
As $\lambda\rightarrow -\infty$, $\ell \rightarrow 0$, so the bias would be infinite and negative if $\beta > 0$.
In reality, there is also the prior opinion which makes the posterior belief behave properly, but whose effect diminishes as $n\rightarrow\infty$, so the bias becomes increasingly negative as
the number of observations increases. Thus such extreme discounting does not lead to a stable bias $\lambda$ as sample size increases.
\subsection{The `sweet spot' model}
When discounting is not directional, there is no bias, but the subjective and true variance of the estimator $\hat\lambda$ change.

Consider a subjective distribution of $X$ that is a t-distribution with $\nu=1/\beta$ degrees of freedom. Then the log-likelihood is
\[\ell(\lambda)=-(1+\beta)\sum_{i=1}^n \frac{\ln(1+\beta(X_i-\lambda)^2/\sigma^2)}{2\beta}.\]
As $\beta\rightarrow 0$ we regain the normal log-likelihood function.

Asymptotically 
\[\ell(\lambda)/n \rightarrow (1+\beta)\text{E}\{\frac{-\ln(1+\beta(X-\lambda)^2/\sigma^2}{2\beta}\},\]
and the peak occurs at
\[\text{E}\{\frac{X-\lambda}{1+\beta(X-\lambda)^2/\sigma^2}\}=0.\]
The only solution is $\lambda=0$, as expected.

The computations of subjective and true variances require integrals $I_1(\beta), I_2(\beta)$,
where
\[I_n(\beta)=\frac{1}{\sqrt{2\pi}}\int_{-\infty}^\infty \frac{\exp(-x^2/2)\dee x}{(1+\beta x^2)^n}.\]
These integrals can be expressed analytically and are derived in the appendix.
Given these functions, the variance of the posterior distribution  $v$ is given by
\[v=-(1/n)\{\partial^2\text{E}(\ell)/\partial\lambda^2|_{\lambda=0}\}^{-1}= \frac{\sigma^2/n}{(1+\beta)(2I_2-I_1)}.\]

\begin{figure}
\centering
\makebox{\includegraphics{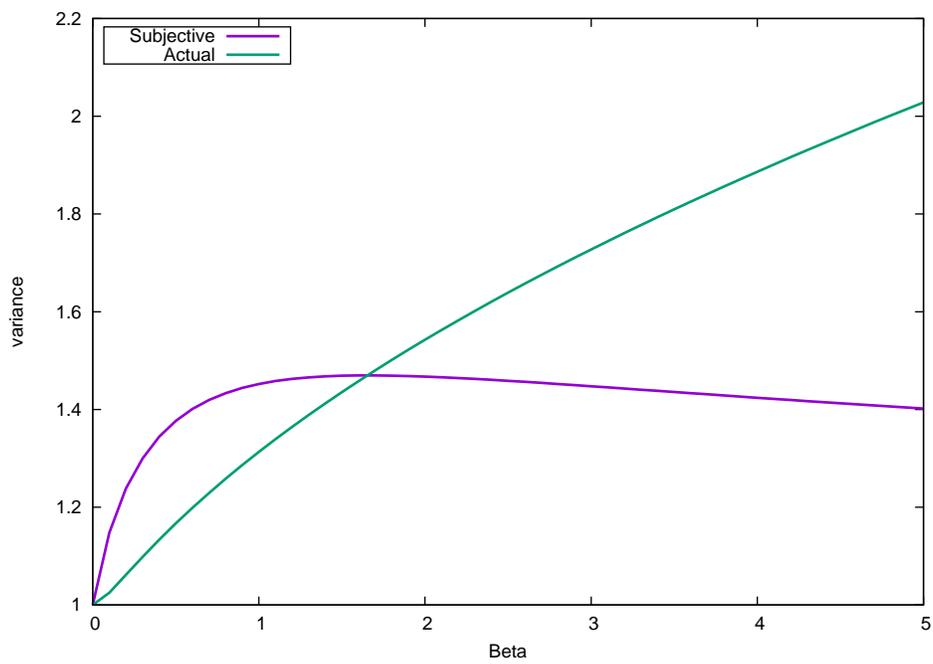}}
\caption{\label{fig3}Subjective and actual asymptotic variance for the posterior distribution mean for the sweet-spot model,
as a function of the amount of bias $\beta$.}
\end{figure}

From (\ref{eq:var}), we have that
\[\text{var}(\hat{\lambda})=(\sigma^2/n)\frac{I_1-I_2}{\beta (2I_2-I_1)^2}.\]
Figure \ref{fig3} shows the subjective and objective variances as a function of the biasing coefficient $\beta$. It can be seen that for large $\beta$, the subjective variance declines whereas the objective variance exceeds it and increases with $\beta$.
\subsection{Constant variance model}
Hence the variance is scaled by $\gamma > 1$ if $X > \lambda$, else scaled by $\beta < 1$. We can take $\beta=1$ so that the variance is not altered for $X < \lambda$, or $\beta=1/\gamma$.
Hence
\[\text{E}(\ell)/n \rightarrow \left\{\begin{array}{ll}
\text{E}\{-(X-\lambda)^2/2\sigma^2\beta\} & \mbox{if $X \le \lambda$} \\
\text{E}\{-(X-\lambda)^2/2\sigma^2\gamma\} & \mbox{if $X > \lambda,$ }
\end{array}
\right. \]
where constants have been discarded.
Differentiating, we have that
\[n^{-1}\text{E}(\partial \ell/\partial \lambda)=(1/\gamma-1/\beta)\frac{\exp(-\lambda^2/2\sigma^2)}{\sqrt{2\pi\sigma^2}}-(\lambda/\sigma^2)\{\frac{\Phi(-\lambda/\sigma)}{\gamma}+\frac{\Phi(\lambda/\sigma)}{\beta}\},\]
whee $\Phi$ is the normal distribution function, and setting the expectation to zero gives an equation for the asymptotic value of $\lambda$.
Also,
\[n^{-1}\text{E}(\partial^2\ell/\partial\lambda^2)=-\sigma^{-2}\{\frac{\Phi(\lambda/\sigma)}{\beta}+\frac{\Phi(-\lambda/\sigma)}{\gamma}\}.\]
There is a solution with $\lambda < 0$ for $\gamma > \beta$.

The most efficient method to compute $\lambda$ is to use Newton-Raphson iteration
\[\lambda_{n+1}=\lambda_n-\frac{\text{E}(\partial \ell/\partial \lambda)}{\text{E}(\partial^2 \ell/\partial\lambda^2)|_{\lambda=\lambda_n}}.\]
This typically converges in 5 or 6 iterations, starting from $\lambda=0$. Since the equation $\text{E}(\partial \ell/\partial \lambda)=0$ contains only the ratio $\lambda/\sigma$, it follows that $\lambda \propto \sigma$.

The variance of the posterior distribution is
\[v=\frac{\sigma^2/n}{\frac{\Phi(\lambda/\sigma)}{\beta}+\frac{\Phi(-\lambda/\sigma)}{\gamma}  }.\]
This can be greater or smaller than $\sigma^2/n$. For $\gamma > \beta$ it is straightforward to derive lower and upper bounds $v > (\sigma^2/n)/2\gamma$,
$v < 2\gamma(\sigma^2/n)$.

The true variance of $\hat\lambda$ can be found from $\partial\ell/\partial\lambda=0$ by applying (\ref{eq:var}). This yields
\[\hat{\lambda}-\lambda=n^{-1}\frac{\sum_{i=0}^{n_1} (X_i-\lambda)/\beta+\sum_{j=1}^{n_2} (Y_j-\lambda)/\gamma}{n_1/\beta+n_2/\gamma},\]
where the $X_i < \lambda$ and $Y_j > \lambda$.
After some manipulation and simplification, this yields
\[\text{var}(\hat{\lambda})=n^{-1}\frac{\{\sigma^2(\Phi(\lambda/\sigma)/\beta^2+\Phi(-\lambda/\sigma)/\gamma^2\}-\lambda^2/\beta\gamma}{\{\Phi(\lambda/\sigma)/\beta+\Phi(-\lambda/\sigma)/\gamma\}^2}.\]
This reduces to $\sigma^2/n$ when $\beta=\gamma$.

The influence function is derivable as
\[n\Delta\hat{\lambda}=\frac{Y-\lambda}{(\gamma/\beta)\Phi(\lambda/\sigma)+\Phi(-\lambda/\sigma)}.\]
This is simply a straight line.

\subsection{Log-gamma distribution model}
Here the subjective distribution for $X$ is log-gamma (see \eg Johnson {\em et al} 1995), so that the log-likelihood is
\begin{equation}\ell=\sum_{i=1}^n \{1-\exp(-\beta(X_i-\lambda)/\sigma)-(\beta/\sigma)(X_i-\lambda),\label{eq:gammaell}\end{equation}
and tends to
\[\ell/n \rightarrow \text{E}\{\frac{1-\exp(-\beta (X-\lambda)/\sigma)-\beta (X-\lambda)/\sigma}{\beta^2}\},\]
ignoring constants. As $\beta \rightarrow 0$ this tends to the correct normal distribution.
Setting $\partial \ell/\partial\lambda=0$ gives $\lambda=-\beta\sigma/2$.
From (\ref{eq:v}) the variance $v=\sigma^2/n$, and is unchanged from the variance that results from using the correct likelihood with $\beta=0$.

From  (\ref{eq:var}) and  using (\ref{eq:ident1}), we have that the true variance is
\[\text{var}(\hat{\lambda})=\frac{\sigma^2}{n}\frac{\exp(\beta^2)-1}{\beta^2},\]

From (\ref{eq:approxlam}), the influence can be derived as
\[n\Delta\hat{\lambda}=(\sigma/\beta)\{1-\exp(-\beta (X-\lambda)/\sigma)\}.\]
Unlike the exponential model case, the influence here always increases with $X$, although sub-linearly. Figure \ref{fig2} shows the function.

\section{Attitude polarization}
Consider the probability that the centre of location is less than a threshold $L$, \ie $\lambda < L$.
Writing $\hat{\lambda}=\lambda+Z\sigma_{\lambda}$, where $\sigma_\lambda$ is the standard deviation of $\hat\lambda$ and $Z$ a standard normal random variable, 
this probability is
\[p=\Phi\{(\hat{\lambda}-L)/\sigma_\lambda\}.\]
As $\sigma_\lambda \propto n^{-1/2}$, $p$ will tend to 0 or 1 as $n\rightarrow\infty$, depending on the sign of $\lambda-L$.
Hence opinion will become increasingly polarized after more and more evidence is observed.
\section{Classification of models}
Several models were explored. In general, with directional discounting whatever the model the bias of the mean of the posterior distribution tends to a small multiple of the standard deviation $\sigma$ of the observations,
and the variance of the posterior distribution decreases proportionally as $\sigma^2/n$. It may be larger or smaller than $\sigma^2/n$, but the decrease means that attitude polarization can occur.
An optimist who tends to believe the high sales figures and discount the low ones as aberrations will become increasingly certain that the figures are high, 
whereas a pessimist will become increasingly certain of the opposite conclusion.

The following properties can be distinguished:
\begin{enumerate}
\item The bias tends to an asymptotic limit in the models given here, but if discounting is strong enough, as in the relative exponential model, it does not.  Hence models can be distinguished by whether or not 
there is an asymptotic limit to bias.

\item The perceived variance is larger in one direction, but may or may not be smaller than the true variance of the observations in the other direction.
For the constant variance model with $\beta=1$ low observations are not unduly prized. However, for the exponential and log-gamma models they are.
This also distinguishes models.

\item For some models such as the constant variance model and the log-gamma model, there is a subjective distribution of observations,
whereas for others such as the exponential model, there is not. Models without a subjective distribution such as the exponential bias model can give an asymptotic fixed bias however,
as long as the expectation of the log-likelihood function is properly behaved.

\item The discounting may be such that higher observations above the mean always have more influence than lower ones, so that $\partial\Delta\hat{\lambda}/\partial X > 0$,
or so extreme that eventually higher observations have less influence than lower ones. Figure \ref{fig2} illustrates the two types of behaviour.
This property seems the best way to order models, based on how much the influence function departs from a straight line. Thus in increasing order of severity of bias, we would have:
the constant variance model, the log-gamma model, the exponential model, followed by models without an asymptotic limit to bias.
\end{enumerate}

\section{Primacy}

Earlier observations have more weight. This could be modelled by taking the perceived variance as $\sigma^2 (\sigma^2/v)^\xi$, where if $\xi > 0$, the variance is smaller 
when the prior distribution has high variance. As observations accumulate, $v$ decreases and the perceived variance increases.

Given a normally-distributed random variable, the variance of the posterior distribution $v^\prime$ is given by
\[\frac{1}{v^\prime}=\frac{1}{v}+\frac{1}{\sigma^2 (\sigma^2/v)^\xi}.\]
We can see how this updating scheme behaves asymptotically when $v$ is small Then if $u=1/v$, we have approximately that
\[\dee u/dt=u^{-\xi}/\sigma^{2+2\xi},\]
where $t$, time, is the number of observations. Hence after $n$ observations, 
\[v=(\sigma^{2\xi+2}/(\xi+1)n)^{1/(\xi+1)},\]
showing that the posterior variance decreases as $n^{-1/(\xi+1)}$. As $\xi$ increases from zero, the decrease is slower. Hence because initial observations have more weight and later ones have less weight,
posterior variance is slow to reduce.

\section{Conclusions}
Some statistical models of confirmation or myside bias have been presented. Most of the models cover the case where observations are normally distributed,
and observations to one side of the believed value are discounted to some extent, while those on the other side may be prized (given unduly high weight).
Such models embody the persistence of prior belief, leading eventually to a posterior mean biased by the order of the standard deviation of the distribution of observations.
The variance of the posterior distribution decreases as the reciprocal of sample size, so those believing in high or low results will polarize opinions, each becoming more certain of their
opposite beliefs as evidence accumulates. The primacy effect can also be accommodated in this framework, where earlier evidence carries more weight than does later evidence.
This could occur if the weight given to an observation is taken to depend on the extent to which it decreases one's ignorance.

This type of model can be applied where rather than an opinion about a continuous quantity such as blood pressure, we have belief in a proposition,
such as that anthropogenic climate change is occurring. Here we imagine an underlying observation, and the probability that the proposition is true is the posterior probability 
to one side of a fixed point $L$. Alternatively, one could use the beta distribution model presented earlier.

This type of modelling would be useful for psychological research into confirmation bias and for assessing its prevalence in an organization.
It would be necessary to elicit belief, \eg in a proposition as evidence accumulates. It would then be possible to fit the models given here, and to estimate
the parameters. However, we may still be far from such research, and the value of this work may be chiefly to stimulate fruitful debate about
the detailed way in which confirmation bias operates.

\section*{Appendix: derivation of integrals for the `sweet-spot' model}
The integrals $I_1, I_2$ are derived, where
\[I_n(\beta)=\frac{1}{\sqrt{2\pi}}\int_{-\infty}^\infty \frac{\exp(-x^2/2)\dee x}{(1+\beta x^2)^n}.\]
These are needed for computing subjective and true variances in the `sweet-spot' model.

The integral $I_1$ is given in Gradshteyn and Ryzhik (2015) as result 3.466 (1), albeit in different notation.
On changing variable, we have
\[I_n(\alpha)=\frac{\sqrt{\alpha}}{\sqrt{2\pi}}\int_{-\infty}^\infty \frac{\exp(-\alpha y^2/2)\dee y}{(1+y^2)^n},\]
where $\alpha=1/\beta$.
Write 
\[J_n(\alpha)=\frac{1}{\sqrt{2\pi}}\int_{-\infty}^\infty \frac{\exp(-\alpha(1+y^2/2))\dee y}{(1+y^2)^n},\]
and differentiate with respect to $\alpha$ to obtain $\dee J_1/\dee \alpha=-\exp(-\alpha/2)/(2\sqrt{\alpha})$.
Integrating, 
\[J_1(\alpha)=-\int_0^\alpha \frac{\exp(-z/2)\dee z}{2\sqrt{z}}+J_1(0).\]
Using $J_1(0)=\sqrt{\pi/2}$ and evaluating the integral, we obtain $J_1$ and finally $I_1$ as
\[I_1=\frac{\sqrt{2\pi}}{\sqrt{\beta}}\Phi(-\sqrt{1/\beta})\exp(1/2\beta).\]
The asymptotic expansion of $\Phi$ shows that $I_1 \rightarrow 1$ as $\beta\rightarrow 0$ as it must.

The integral $I_2$ is not given in Gradshteyn and Ryzhik. We have $\dee J_2/\dee\alpha=-\frac{1}{2}J_1=\sqrt{2\pi}\Phi(-\sqrt{\alpha})$.
From this,
\[J_2(\alpha)=\frac{\sqrt{\pi}}{\sqrt{2}}\int_0^\alpha \Phi(-\sqrt{x})\dee x+J_2(0).\]
Since $J_2(0)=\frac{\sqrt{\pi}}{2\sqrt{2}}$, $J_2(\alpha)$ can be evaluated by twice integrating by parts.
Finally,
\[I_2=\frac{1}{2\beta}-\frac{\sqrt{\pi}}{\sqrt{2}}\frac{(1-\beta)}{\beta^{3/2}}\exp(1/2\beta)\Phi(-1/\sqrt{\beta}).\]
The asymptotic expansion of $\Phi$ to second order shows that $I_2(0)=1$ as it must.

\end{document}